\documentclass{emulateapj}
\usepackage{apjfonts}
\journalinfo{The Astrophysical Journal, 2010, in press}

\shorttitle{AN APPROXIMATE CORONAL HEATING RATE}
\shortauthors{CRANMER}

\begin{document}

\title{An Efficient Approximation of the Coronal Heating Rate
for Use in Global Sun-Heliosphere Simulations}

\author{Steven R. Cranmer}
\affil{Harvard-Smithsonian Center for Astrophysics,
60 Garden Street, Cambridge, MA 02138, USA}
\email{scranmer@cfa.harvard.edu}

\begin{abstract}
The origins of the hot solar corona and the supersonically expanding
solar wind are still the subject of debate.
A key obstacle in the way of producing realistic simulations of
the Sun-heliosphere system is the lack of a physically motivated
way of specifying the coronal heating rate.
Recent one-dimensional models have been found to reproduce many
observed features of the solar wind by assuming the energy
comes from Alfv\'{e}n waves that are partially reflected, then
dissipated by magnetohydrodynamic turbulence.
However, the nonlocal physics of wave reflection has made it
difficult to apply these processes to more sophisticated
(three-dimensional) models.
This paper presents a set of robust approximations to the
solutions of the linear Alfv\'{e}n wave reflection equations.
A key ingredient to the turbulent heating rate is the
ratio of inward to outward wave power, and the approximations
developed here allow this to be written explicitly in terms of
local plasma properties at any given location.
The coronal heating also depends on the frequency spectrum of
Alfv\'{e}n waves in the open-field corona, which has not yet
been measured directly.
A model-based assumption is used here for the spectrum, but
the results of future measurements can be incorporated easily.
The resulting expression for the coronal heating rate is
self-contained, computationally efficient, and applicable
directly to global models of the corona and heliosphere.
This paper tests and validates the approximations by comparing the
results to exact solutions of the wave transport equations in
several cases relevant to the fast and slow solar wind.
\end{abstract}


\section{Introduction}

The hot, ionized outer atmosphere of the Sun is a unique laboratory
for the study of magnetohydrodynamics (MHD) and plasma physics.
Despite more than a half-century of study (Parker 1958), the basic
physical processes responsible for heating the million-degree
solar corona and accelerating the solar wind are still not known.
Identification of these processes is important not only for
understanding the origins and impacts of space weather
(e.g., Feynman \& Gabriel 2000; Eastwood 2008),
but also for establishing a baseline of knowledge about a
well-resolved star that is directly relevant to other
astrophysical systems.

In recent years, two general paradigms have emerged as attempts
to address how both fast and slow solar wind streams are heated
and accelerated.
In the {\em wave/turbulence-driven} (WTD) class of models,
it is generally assumed that the convection-driven jostling of
magnetic flux tubes in the photosphere drives wave-like fluctuations
that propagate up into the extended corona.
These waves (usually Alfv\'{e}n waves) are proposed to partially
reflect back down toward the Sun, develop into MHD turbulence, and
heat the plasma by their gradual dissipation (see, e.g.,
Hollweg 1986; Wang \& Sheeley 1991; Matthaeus et al.\  1999;
Suzuki \& Inutsuka 2006; Cranmer et al.\  2007).
In the {\em reconnection/loop-opening} (RLO) class of models,
the flux tubes feeding the solar wind are assumed to be influenced
by impulsive bursts of mass, momentum, and energy deposition in
the low atmosphere.
This energy is usually assumed to come from magnetic reconnection
between closed, loop-like magnetic flux systems and the open
flux tubes that connect to the solar wind (see, e.g.,
Axford \& McKenzie 1992; Fisk et al.\  1999;
Schwadron \& McComas 2003).

Determining whether the WTD or RLO paradigm---or some combination
of the two---is the dominant cause of global solar wind
variability is a key prerequisite to building physically
realistic models of the heliosphere.
One way to make progress is to include either WTD or RLO processes
in existing three-dimensional numerical simulations and compare the
results with measurements.
The main goal of this paper is to provide a new set of tools
that allows the incorporation of WTD physics into simulations of the
Sun-heliosphere system.
Many of the widely-applied three-dimensional modeling codes have
used relatively simple empirical prescriptions for coronal heating
in the energy conservation equations (Riley et al.\  2001, 2006;
Roussev et al.\  2003; T\'{o}th et al.\  2005;
Usmanov \& Goldstein 2006; Feng et al.\  2007;
Lionello et al.\  2009; Sokolov et al.\  2009;
Nakamizo et al.\  2009; Schmit et al.\  2009).
It would also be beneficial to apply more realistic heating rates
to more focused studies of solar wind expansion, such as
two-dimensional axisymmetric models of coronal streamers
(e.g., V\'{a}squez et al.\  2003; Endeve et al.\  2004).

The starting point for this work is an existing model of
Alfv\'{e}n wave reflection and dissipative heating
(Cranmer \& van Ballegooijen 2005).
In that model, an observationally constrained set of plasma
parameters in a polar coronal hole was specified as a
time-steady background state on which the properties of
waves and turbulence were computed.
Subsequently, the same phenomenological wave physics was
inserted into a self-consistent solution of the equations of
mass, momentum, and energy conservation (Cranmer et al.\  2007).
The only input ``free parameters'' to these models of coronal
heating and solar wind acceleration were the photospheric
lower boundary conditions (for the waves) and the radial
dependence of the background magnetic field.
For a single choice for the lower boundary condition, these
models produced a realistic variation of fast and slow solar
wind conditions by varying only the coronal magnetic field
(see also Cranmer 2009).
There has been a great deal of other recent work done to improve
our understanding of Alfv\'{e}n wave reflection and MHD turbulence
as a source of coronal heating (e.g.,
Verdini et al.\  2005, 2009; Verdini \& Velli 2007;
Rappazzo et al.\  2007, 2008; Chandran \& Hollweg 2009).

Despite this progress, it has been difficult to apply the results
of these focused studies to more comprehensive models of
the Sun-heliosphere system.
Determining even the most basic ingredients of the theoretical
coronal heating rate requires the solution of an additional set of
differential equations for the rate of Alfv\'{e}n wave reflection.
These equations depend on the wave frequency, and they are
inherently nonlocal in their dependence on the plasma
parameters along a given flux tube.
In other words, the amount of wave reflection at any given
location in the corona appears to depend on an integration
over distance, and not just on the local plasma properties.
Thus, in order to compute the coronal heating from
reflection-driven turbulence, it has been necessary to solve a
set of differential equations for {\em each frequency} in a
continuous power spectrum, for {\em each flux tube} of
interest that fills three-dimensional space, and for
{\em each time step} of a simulation.

This paper presents a set of approximations that allows the
rate of Alfv\'{e}n wave reflection to be computed without the
need for computationally intensive solutions of differential
equations.
The new approximations are completely {\em local} in nature, in
that they depend only on the plasma parameters at the location in
the corona at which the coronal heating rate is to be computed.
It is hoped that these approximations will speed up the
calculation of the coronal heating rate by orders of magnitude
in comparison to earlier studies.
Section 2 discusses the relevant equations and approximations.
Section 3 compares the exact (numerically integrated) reflection
coefficients and heating rates with those computed from the
approximations.
Section 4 describes a FORTRAN subroutine that has been
developed to implement these approximations.
The code is included with this paper as online-only material.
Finally, Section 5 concludes this paper with a brief summary
of the major results and a discussion of additional physical
processes that can be included to improve the modeling of
the corona and solar wind.

\section{Alfv\'{e}n Wave Reflection}

This paper considers the one-dimensional variation of plasma
parameters along a magnetic flux tube that is rooted in the solar
photosphere and extends into interplanetary space.
The general assumption will be that the corona and solar wind
are in a state of steady (i.e., time independent) expansion.
However, the heating rates discussed below may be valid under
time-variable conditions as well.
Throughout this section, the numerical examples are taken from
an observationally constrained model of a polar coronal hole at
solar minimum (Cranmer \& van Ballegooijen 2005).

\subsection{The Linear Non-WKB Reflection Problem}

Alfv\'{e}n waves are modeled here as linear, incompressible,
and transverse fluctuations that propagate along a magnetic
flux tube with background field strength $B_{0}$.
The wave perturbations in velocity and magnetic field are
denoted $v_{\perp}$ and $B_{\perp}$.
The assumption of linearity is consistent with the limiting
case $| B_{\perp} / B_{0} | \ll 1$.
In general, the perturbed wave properties are complex quantities
that vary as a function of both time $t$ and heliocentric radius $r$.
The conservation equations written below implicitly assume linear
polarization of the waves along a single transverse dimension.
However, this assumption does not limit the applicability of
the resulting wave amplitudes to other polarization states
(see, e.g., Heinemann \& Olbert 1980).

It is convenient to express the wave amplitudes in terms of
Els\"{a}sser (1950) variables, which are defined in velocity
units as
\begin{equation}
  z_{\pm} \, \equiv \, v_{\perp} \pm
  \frac{B_{\perp}}{\sqrt{4 \pi \rho}}  \,\, ,
  \label{eq:elsdefine}
\end{equation}
where $\rho$ is the local mass density, $z_{-}$ represents
outward propagating waves, and $z_{+}$ represents inward
propagating waves.
(This is a convention-dependent assignment; other papers often
use other definitions.)
In a frame of reference flowing with the solar wind, these
oscillations propagate up and down along the field lines with
phase and group speeds equal to the local Alfv\'{e}n speed
$V_{\rm A} = B_{0} / (4\pi\rho)^{1/2}$.

If the waves are propagating in only one direction along the field,
the radial variation of their amplitude and phase can be described
straightforwardly by defining a local wavenumber and utilizing
the concept of {\em wave action conservation}
(e.g., Jacques 1977).
This limiting case is often described in terms of the WKB
(Wentzel, Kramers, Brillouin) approximation.
However, the more general case of counterpropagating
Alfv\'{e}n waves (i.e., a superposition of both Els\"{a}sser
components) tends to require a non-WKB treatment.
In this case, the radial evolution of the oscillation profile 
can no longer be expressed with a local wavenumber.
It has been known for some time that a spatially varying
Alfv\'{e}n speed allows for gradual linear reflection
(Ferraro \& Plumpton 1958).
This problem has been studied extensively in the context of solar
and stellar winds (e.g., Hollweg 1978, 1981, 1990; Wentzel 1978;
Heinemann \& Olbert 1980; An et al.\  1990; Barkhudarov 1991;
Velli 1993; Krogulec et al.\  1994; MacGregor \& Charbonneau 1994;
Orlando et al.\  1996; Laitinen 2005; Verdini et al.\  2005).

For the solar models considered here, the magnitude of outward
propagating waves always remains larger than the magnitude of
inward (i.e., reflected) waves, and thus $|z_{+}| / |z_{-}| < 1$.
At large heights in the corona and solar wind, it is often the
case (for some frequencies) that the reflection is very inefficient,
and $|z_{+}| / |z_{-}| \ll 1$.
It's important to note, however, that the sharp transition region
(TR) between the chromosphere and corona can act as an efficient
``reflection barrier'' to Alfv\'{e}n waves.
Thus, in the photosphere and chromosphere, the reflection can be
considered nearly complete ($|z_{+}| / |z_{-}| \approx 1$) and
the fluctuations are similar in character to standing waves.

The incompressible first-order conservation equations for
mass and momentum, as well as the magnetic induction equation,
can be transformed into a pair of wave transport equations,
\begin{equation}
  \frac{\partial z_{\pm}}{\partial t} + ( u \mp V_{A} )
  \frac{\partial z_{\pm}}{\partial r} \, = \, ( u \pm V_{A} )
  \left( \frac{z_{\pm}}{4 H_{\rm D}} +
  \frac{z_{\mp}}{2 H_{\rm A}} \right)
  \label{eq:zpm}
\end{equation}
where $u$ is the solar wind speed and the signed scale
heights are defined as
$H_{\rm D} \equiv \rho / (\partial \rho / \partial r)$ and
$H_{\rm A} \equiv V_{\rm A} / (\partial V_{\rm A} / \partial r)$.
Various alternate ways of writing Equation (\ref{eq:zpm}) are
described in Appendix B of Cranmer \& van Ballegooijen (2005).
The phenomenon of gradual linear reflection arises because of the
presence of the $z_{\mp}/H_{\rm A}$ term on the right-hand side.
This produces coupling between the two Els\"{a}sser variables.

Equation (\ref{eq:zpm}) does not contain any terms that describe
the nonlinear damping of Alfv\'{e}n waves.
Cranmer \& van Ballegooijen (2005) found that this damping does
not strongly affect the wave amplitudes in the corona, but it
may be an important effect at larger distances in the heliosphere.
Thus, it appears justifiable to separate the problem of non-WKB
reflection from that of the nonlinear damping and coronal heating.
This is what is done in this paper.
Alternately, Chandran \& Hollweg (2009) presented a set of
approximations of non-WKB reflection in which the nonlinear
damping was included explicitly in the transport equations.
It remains to be seen whether a combination of the approximations
developed in this paper with those of Chandran \& Hollweg (2009)
will yield an improved description of the overall wave transport
and dissipation.

If the complete radial and time dependence of $z_{+}$ and
$z_{-}$ were known for a given flux tube in the solar wind, it
would be possible to compute the turbulent heating rate
(see Section 3 below).
The exact solution of Equation (\ref{eq:zpm}), however,
traditionally requires either numerical relaxation or direct
integration up and down along the flux tube, starting at the
Alfv\'{e}n critical point.
This has made it difficult to incorporate an accurate description
of reflection-driven turbulence in three-dimensional
Sun-heliosphere simulations.

Barkhudarov (1991) presented a dimensionless version of the
transport equations in which the Els\"{a}sser variables are
expressed as
\begin{equation}
  z_{\pm}(r,t) \, = \, G_{\pm}(r) \, \exp \left\{ i \left[
  \Gamma_{\pm}(r) + \omega t \right] \right\}
\end{equation}
where the angular frequency $\omega$ (expressed in rad s$^{-1}$)
is a real constant.
The amplitudes $G_{\pm}(r)$ and angular phases $\Gamma_{\pm}(r)$
are real functions of distance along the flux tube.
In order to determine the degree of non-WKB wave reflection, one
can solve for two dimensionless quantities.
First, a scaled ratio of the two amplitudes can be defined as
\begin{equation}
  \Psi \, = \, \left( \frac{u - V_{A}}{u + V_{A}} \right)
  \frac{G_{+}}{G_{-}} \,\, .
\end{equation}
Second, the angular phase shift between the inward and outward
wave trains is defined as $\Gamma = \Gamma_{+} - \Gamma_{-}$.
Following the terminology of Cranmer et al.\  (2007), one can
also define an effective frequency-dependent ``reflection
coefficient'' as ${\cal R} = |z_{+}| / |z_{-}| = G_{+}/G_{-}$.
This is the primary quantity that the approximations of this paper
are designed to estimate.

Barkhudarov (1991) discussed how the transport equations can be
transformed into dimensionless conservation equations for the
two non-WKB quantities $\Psi$ and $\Gamma$.
These equations are
\begin{equation}
  \frac{d\Psi}{dr} \, = \, \frac{(\Psi^{2} - 1) \cos\Gamma}
  {2 H_{\rm A}}
  \label{eq:dPsi}
\end{equation}
\begin{equation}
  \frac{d\Gamma}{dr} \, = \, \frac{(\Psi^{2} + 1) \sin\Gamma}
  {2 H_{\rm A} \Psi} - \frac{2 \omega V_{\rm A}}{u^{2} -
  V_{\rm A}^{2}}  \,\, .
  \label{eq:dGam}
\end{equation}
Although Barkhudarov's (1991) derivations were limited to the case
of pure spherical expansion (i.e., $B_{0} \propto r^{2})$, the
above equations and definitions have been shown to be valid for
an arbitrary flux-tube expansion factor (see
Cranmer \& van Ballegooijen 2005).
If the phase shift $\Gamma$ is known, one can use 
Barkhudarov's closed-form solution to Equation (\ref{eq:dPsi}) to
solve for
\begin{equation}
  \Psi \, = \, \frac{1 - e^{2W}}{1 + e^{2W}}  \,\, ,
\end{equation}
where
\begin{equation}
  W(r) \, = \, \int_{r_{0}}^{r} dr \,\,
  \frac{\cos\Gamma}{2 H_{\rm A}}  \,\, .
  \label{eq:Wdef}
\end{equation}
In other words, if it is possible to obtain an expression for $W$
in terms of radius, wave frequency, and the local plasma properties,
then one can straightforwardly determine $\Psi$ and thus ${\cal R}$.

The Alfv\'{e}n critical point (at which $u = V_{\rm A}$) is a
singular point of the transport equations.
This critical radius is denoted as $r_0$.
When $r = r_{0}$, the wind speed and Alfv\'{e}n speed both have
the identical value $V_0$.
The ratio $\Psi$ is zero at this critical point, it is negative
where $r < r_0$, and it is positive where $r > r_0$.
Over the full range of distances, $| \Psi | < 1$.
Barkhudarov (1991) derived the following constraint on the
phase shift $\Gamma$ at the singular point,
\begin{equation}
  \tan \Gamma_{0} \, = \, \frac{\omega}{\mu_0}  \,\, ,
  \label{eq:tanG0}
\end{equation}
where
\begin{equation}
  \mu \, = \, \frac{du}{dr} - \frac{dV_{\rm A}}{dr}  \,\, .
  \label{eq:mudef}
\end{equation}
It is usually the case that the Alfv\'{e}n speed gradient
is the dominant contributor in the definition of $\mu$, and one
can often safely ignore the $du/dr$ term above.
The quantity $\mu_0$ is the value of $\mu$ at the Alfv\'{e}n
critical point, and this (usually positive) quantity acts as
an effective ``cutoff frequency'' for wave reflection (see,
e.g., Musielak et al.\  1989).
Waves having frequencies much lower than $\mu_0$ are strongly
reflected.
Waves having frequencies much higher than $\mu_0$ are reflected
very weakly and behave similarly to WKB-like oscillations that
obey wave action conservation.
An application of L'H\^{o}pital's rule gives a concise expression
for the reflection coefficient at the critical point, which is
\begin{equation}
  {\cal R}_{0} \, = \, \frac{| dV_{\rm A}/dr |_{0}}
  {\sqrt{\omega^{2} + \mu_{0}^{2}}}  \,\, .
\end{equation}

\subsection{Reflection in the Zero Frequency Limit}

In order to find approximate expressions for the amount of
non-WKB wave reflection at heights other than the Alfv\'{e}n
critical point, it is useful to examine the solutions to the
dimensionless wave transport equations in various liming cases.
Numerical models show that in the limit of very low wave
frequency (i.e., $\omega \rightarrow 0$), the angular
phase shift $\Gamma$ also approaches zero over all radii.
Thus, since $\cos\Gamma \approx 1$ everywhere,
Equation (\ref{eq:Wdef}) reduces to
\begin{equation}
  W \, = \, \frac{1}{2} \int_{r_{0}}^{r} dr \,\,
  \frac{d \, \ln V_{\rm A}}{dr} \, = \,
  \frac{1}{2} \ln \left( \frac{V_{\rm A}}{V_0} \right) \,\, .
  \label{eq:Wzero}
\end{equation}
This particularly simple solution leads to a closed-form
expression for the reflection coefficient, with
\begin{equation}
  {\cal R}_{\rm zero} \, = \, \left( 
  \frac{u+V_{\rm A}}{u-V_{\rm A}} \right)
  \left( \frac{V_{0}-V_{\rm A}}{V_{0}+V_{\rm A}} \right) \,\, .
  \label{eq:Rzero}
\end{equation}
This form of the reflection coefficient has been shown to agree
well with the numerical solutions of
Cranmer \& van Ballegooijen (2005) and Cranmer et al.\  (2007)
at the lowest modeled frequencies of $\sim 10^{-6}$ Hz.
Thus, if the majority of the Alfv\'{e}n wave power in the corona
and solar wind is at low enough frequencies ($\omega \ll \mu_0$),
then Equation (\ref{eq:Rzero}) provides the ratio of
counterpropagating wave amplitudes as a function of only local
plasma parameters ($u$, $V_{\rm A}$) and the velocity at the
Alfv\'{e}n critical point ($V_0$).

The above zero-frequency limit for the rate of linear reflection
should not be confused with a similarly named ''low-frequency''
approximation used in studies of solar wind turbulence.
A series of MHD scale-separation models has been developed over
the past few decades (e.g., Zhou \& Matthaeus 1990;
Zank et al.\  1996; Matthaeus et al.\  1999, 2004;
Breech et al.\  2005, 2008; Usmanov et al.\  2009)
in which the radial dependence of the power in the $z_{+}$ and
$z_{-}$ modes is computed.
The results are often expressed in terms of the normalized
cross helicity, $\sigma_{c} = (1- {\cal R}^{2})/(1+ {\cal R}^{2})$.
In these models the fluctuation power is assumed to be dominated by
the lowest frequency modes.
Some of these studies explicitly included linear wave reflection
(e.g., Matthaeus et al.\  1999) and some included only other
processes such as large-scale shears, pickup protons in the
outer heliosphere, and turbulent ``dynamic alignment'' (for a
comprehensive summary, see Breech et al.\  2008).

\subsection{Reflection in the Infinite Frequency Limit}

Alfv\'{e}n waves having higher frequencies than described above
(i.e., $\omega \gtrsim \mu_0$) undergo substantially weaker
reflection than in the zero-frequency limit.
In the very high frequency limit of $\omega \gg \mu_0$, the
numerical models of Barkhudarov (1991) and
Cranmer \& van Ballegooijen (2005) showed that the phase shift
$\Gamma$ approaches an asymptotic value of $-\pi/2$ at all radii.
In other words, $\cos\Gamma \rightarrow 0$, and the magnitude
of $W$ becomes very close to zero everywhere, as well.
This also drives $\Psi$ and ${\cal R}$ to small absolute values.
At the Alfv\'{e}n critical point,
\begin{equation}
  \cos\Gamma \, = \, \frac{\mu}{\sqrt{\omega^{2} + \mu^2}}
  \, \approx \, \frac{\mu}{\omega} \, \ll \, 1 \,\, ,
  \label{eq:cGinf}
\end{equation}
where the latter approximation holds for large frequencies
(see Equation (\ref{eq:tanG0})).
It is evident from the numerical models of high-frequency wave
reflection that Equation (\ref{eq:cGinf}) serves reasonably well
as an approximation for the {\em entire radial dependence} of the
magnitude of $\cos\Gamma$, and not merely for its value at $r_0$.

The realization that Equation (\ref{eq:cGinf}) may be valid
over all radii can be used to estimate $W$.
An illustrative example is to examine the properties of the
solar wind at large radii, where $V_{\rm A} \propto r^{-1}$.
This condition is usually satisfied for $r \gtrsim r_0$, and
in some cases it is also a reasonable approximation over a few
solar radii below $r_0$ as well.
Assuming that the Alfv\'{e}n speed gradient dominates the
definition of $\mu$ (i.e., Equation (\ref{eq:mudef})), then
$\mu \approx V_{\rm A}/r \approx V_{0} r_{0} / r^{2}$.
Applying this to Equations (\ref{eq:Wdef}) and (\ref{eq:cGinf}),
it becomes possible to solve for
\begin{equation}
  W \, \approx \, \frac{V_{0} r_{0}}{4\omega} \left(
  \frac{1}{r^2} - \frac{1}{r_{0}^2} \right) \, \approx \,
  \frac{\mu - \mu_{0}}{4\omega}  \,\, .
\end{equation}
It was found that a slightly better approximation is to replace
the factor of $\omega$ above with the full denominator of
Equation (\ref{eq:cGinf}).
Thus, an improved approximate form for $W$ in the limit of
high wave frequencies is
\begin{equation}
  W_{\infty \mu} \, = \, 
  \frac{(\mu - \mu_{0}) \, s}{4 \sqrt{\omega^{2} + \mu^2}}
  \label{eq:Winfmu}
\end{equation}
where $s$ is a dimensionless constant that usually is equal to 1,
but sometimes needs to be set to $-1$ (see below).

It should be noted that Equation (\ref{eq:Winfmu}) provides a
good approximation to the numerical solutions at large radii, but
it begins to fail closer to the Sun, where the Alfv\'{e}n speed
departs from an $r^{-1}$ radial dependence.
One problem with Equation (\ref{eq:Winfmu}) is that the solution for
$W$ changes sign whenever the numerator ($\mu - \mu_0$) changes sign.
Most modeled radial profiles for $V_{\rm A}(r)$ show at least
one local maximum in the corona, and sometimes two or more maxima
(e.g., Cranmer \& van Ballegooijen 2005; Evans et al.\  2008).
However, the numerical solutions for Alfv\'{e}n wave reflection
consistently show that $W > 0$ for $r < r_{0}$, and $W < 0$ for
$r > r_{0}$.
Practically, this can be remedied by setting $s = -1$ in
Equation (\ref{eq:Winfmu}) whenever the condition
$\mu < \mu_{0}$ occurs at radii $r < r_0$.
This is equivalent to taking the absolute value of Equation
(\ref{eq:Winfmu}) everywhere below the Alfv\'{e}n critical point.
Doing this gives values for $W$ (and $\Psi$ and ${\cal R}$) that
are in better agreement with the numerical models.
Nonetheless, this expression still gives rise to unphysical dips
to $W=0$ in the narrow radial zones where $\mu - \mu_0$ changes
sign.
The numerical models do not exhibit $W = 0$ at these locations.

An alternate approximation is to replace the derivatives in the
definition of $\mu$ with a positive-definite expression that
involves only the plasma parameters themselves.
Experimentation with a range of functional forms led to the
definition of
\begin{equation}
  \nu \, = \, \frac{r V_{\rm A}}{(r+R_{\odot})(r-R_{\odot})}
  \label{eq:nu}
\end{equation}
where $R_{\odot}$ is the solar radius, and it can be seen that
$\nu \approx \mu$ for large radii $r \gg R_{\odot}$.
This expression can be substituted for $\mu$ in
Equation (\ref{eq:Winfmu}) to form an alternate definition
for $W$,
\begin{equation}
  W_{\infty \nu} \, = \, 
  \frac{\nu - \nu_{0}}{4 \sqrt{\omega^{2} + \nu^2}} \,\, ,
  \label{eq:Winfnu}
\end{equation}
where $\nu_0$ is the value of $\nu$ at the Alfv\'{e}n critical point.
Figure 1 shows the radial dependence of $\mu$ and $\nu$, as well as
the even simpler approximation $\mu \approx V_{\rm A}/r$ used above,
for the coronal hole model of Cranmer \& van Ballegooijen (2005).
Note that $\nu \approx | \mu |$ nearly everywhere, but $\nu$
remains continuous and positive at the locations in the corona
where $\mu$ changes sign.
The simpler expression $V_{\rm A}/r$ disagrees significantly with
the magnitudes of both $\mu$ and $\nu$ at low heights in the corona.

\begin{figure}
\epsscale{1.09}
\plotone{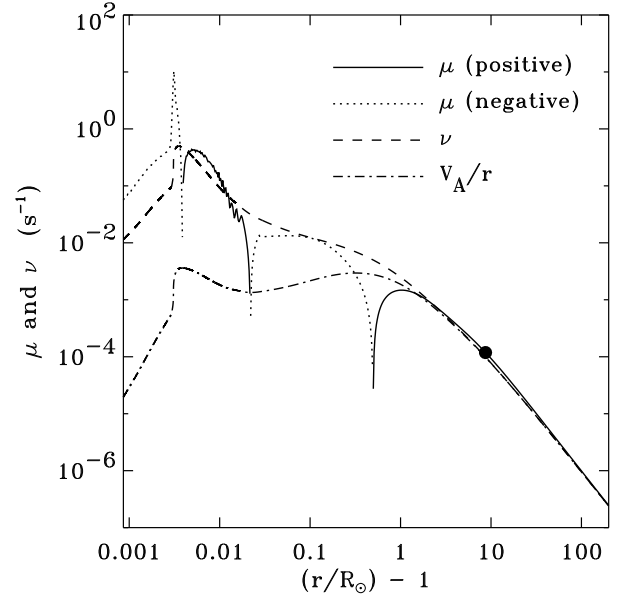}
\caption{Approximations of the Alfv\'{e}n speed gradient for
a polar coronal hole, plotted as a function of height above
the solar photosphere.
The magnitude of $\mu$ is shown in regions where $\mu > 0$
(solid curves) and where $\mu < 0$ (dotted curves).
The positive-definite estimates $\nu$ (dashed curve) and
$V_{\rm A}/r$ (dot-dashed curve) are also shown.  The filled
circle denotes $\mu_0$ at the Alfv\'{e}n critical point.}
\end{figure}

In practice, an even better high-frequency estimate of ${\cal R}$
was found by taking the arithmetic average of the resulting
reflection coefficients that come from Equations (\ref{eq:Winfmu})
and (\ref{eq:Winfnu}).
As can be seen below, there is a moderate amount of ``spikiness''
in the numerically computed reflection coefficients at the radii
where $\mu$ changes sign.
Using only Equation (\ref{eq:Winfmu}) would have overestimated
that effect, and using only Equation (\ref{eq:Winfnu}) would
have underestimated it.

\subsection{Bridging for Intermediate Frequencies}

The two extreme limits of low frequencies (Section 2.2) and
high frequencies (Section 2.3) can be combined to produce a
more complete estimate for the coupled radius and frequency
dependence of the reflection coefficient.
The most robust way of doing this is to first compute the separate
values of ${\cal R}$ for each of the above limiting forms of
$W$, and then combine them in the following way:
\begin{equation}
  {\cal R} \, = \, {\cal R}_{\rm zero}^{1-\epsilon} \left[
  \frac{{\cal R}_{\infty \mu} + {\cal R}_{\infty \nu}}{2}
  \right]^{\epsilon}  \,\, .
  \label{eq:bridgem}
\end{equation}
The dimensionless bridging exponent $\epsilon$ needs to be a
function of both radius and wave frequency.
When $0 < \epsilon < 1$, the above functional form gives a
resulting value of ${\cal R}$ that is intermediate between the
low and high frequency limiting cases.
Some experimentation found the optimal definition
\begin{equation}
  \epsilon \, = \, \frac{1}{1 + (\omega_{0} / \omega)^{2}}
\end{equation}
where $\omega_{0} = (\nu + \nu_0)/2$.
At heights where $\omega$ is much smaller than the local value of
$\omega_0$, the exponent $\epsilon \approx 0$ and
Equation (\ref{eq:bridgem}) is dominated by the low-frequency
limit ${\cal R}_{\rm zero}$.
On the other hand, when $\omega \gg \omega_0$, the exponent
$\epsilon$ approaches 1 and the bridging relation is dominated
by the average of the two alternate forms of the high-frequency
limit for ${\cal R}$.

What is the behavior of the ``bridging frequency'' $\omega_0$?
At small radii (i.e., in the corona) where the local value of
$\nu$ is large compared to $\nu_0$, the bridging frequency
$\omega_0$ is dominated by that local value.
However, at large radii above the critical point, where
$\nu \ll \nu_0$, the bridging frequency is not allowed to
decrease below $\nu_{0}/2$.
This behavior forms a boundary between the high and low
frequency regions (in radius-frequency space) that matches
what is seen in the numerical solutions.

It should be noted that the above estimates are intended to
apply to the corona and the solar wind, and {\em not} to the
photosphere and chromosphere.
For the latter (low-temperature, high-density) regions that
sit below the sharp TR, a relatively safe approximation
would be to simply assign ${\cal R} \approx 1$.
In any case, it is likely that in the chromosphere, other
sources of heating---such as the entropy gain at shocks
formed by the steepening of acoustic waves---are more
important than Alfv\'{e}n waves (e.g., Cranmer et al.\  2007).

\subsection{Estimating Properties of the Alfv\'{e}n Critical Point}

In one-dimensional models of the plasma conditions along a
specified magnetic flux tube, the location of the Alfv\'{e}n
critical point can be found rather easily.
In that case, the numerical values of $V_0$, $\mu_0$, and
$\nu_0$ would also be known.
However, in multi-dimensional MHD simulations, in which the
conservation equations are solved either by discretization or
by spectral methods, it may not be feasible to calculate these
quantities for each point in space and time.
Doing so would require computationally intensive integrations
up and down along individual magnetic field lines, at each
time step.
Since the main purpose of this paper is to eliminate the need for
similar kinds integrations of the non-WKB equations, it would be
advantageous to find approximate ways of computing the properties
of the Alfv\'{e}n critical point---even when all one knows are
the properties at a location far from this point.
Thus, in order to be able to apply the techniques developed in
Sections 2.2--2.4 to as wide a range of models as possible,
this subsection presents an ``optional'' method to estimate
$r_0$ and $V_0$ when these are not known a~priori.

For a steady-state solar wind, the condition of mass flux
conservation demands that the quantity $\rho u / B_{0}$ remain
constant along any flux tube.
Substituting in the definition of the Alfv\'{e}n speed, this
condition implies that the density at the Alfv\'{e}n critical
point is determined uniquely to be
\begin{equation}
  \rho_{0} \, = \, \mbox{constant} \, = \,
  \rho \left( \frac{u}{V_{\rm A}} \right)^{2}  \,\, ,
\end{equation}
where all quantities on the right-hand side are evaluated at any
arbitrary location along the flux tube.
It is interesting that a value for $\rho_0$ can be computed even
if neither $r_0$ nor $V_0$ are known for that flux tube.
The ratio $\rho / \rho_{0}$ is useful as a definitive probe of
whether the current location in the corona or solar wind is
below ($\rho/\rho_{0} > 1$) or above ($\rho/\rho_{0} < 1$) the
Alfv\'{e}n critical point.

In the numerical models of Cranmer \& van Ballegooijen (2005)
and Cranmer et al.\  (2007), the value of $V_0$ is always
intermediate between the instantaneous values of $u$ and
$V_{\rm A}$.\footnote{%
This does not apply to the chromospheric and photospheric
regions below the TR.
At those low heights, both the local values of $u$ and
$V_{\rm A}$ often dip below $V_{0}$.
In these regions, however, the methods outlined in Sections
2.2--2.5 do not need to be used and ${\cal R} = 1$ is not a bad
approximation.}
Thus, a parameterization of the form
\begin{equation}
  V_{0} \, = \, u^{\alpha} V_{\rm A}^{1-\alpha}
  \label{eq:V0est}
\end{equation}
was found to be useful, where $0 < \alpha < 1$.
Some trial-and-error experimentation led to a density-dependent
fit for the exponent, with
\begin{equation}
  \alpha \, \approx \, \frac{1}{1 + 0.3 (\rho / \rho_{0})^{0.25}}
  \,\, .
  \label{eq:alpha}
\end{equation}
Figure 2 shows the radial dependence of the estimated value of $V_0$
for the coronal hole model of Cranmer \& van Ballegooijen (2005).
This approximation gives a roughly constant magnitude for $V_0$
throughout most of the corona and solar wind.
However, there is likely to be room for improvement in the choices
for the numerical constants in Equation (\ref{eq:alpha}).
This parameterization should be tested and refined by comparing
with additional models of the corona and solar wind.
Once $V_0$ is known, a reasonable approximation for the radius of
the Alfv\'{e}n critical point is
\begin{equation}
  r_{0} \, \approx \, \frac{r \, V_{\rm A}}{V_{0}}  \,\, .
\end{equation}
This is an exact expression over the range of heights at which
$V_{\rm A} \propto r^{-1}$, but it appears to also provide a
decent order-of-magnitude estimate for $r_0$ at other heights
as well.
\begin{figure}
\epsscale{1.17}
\plotone{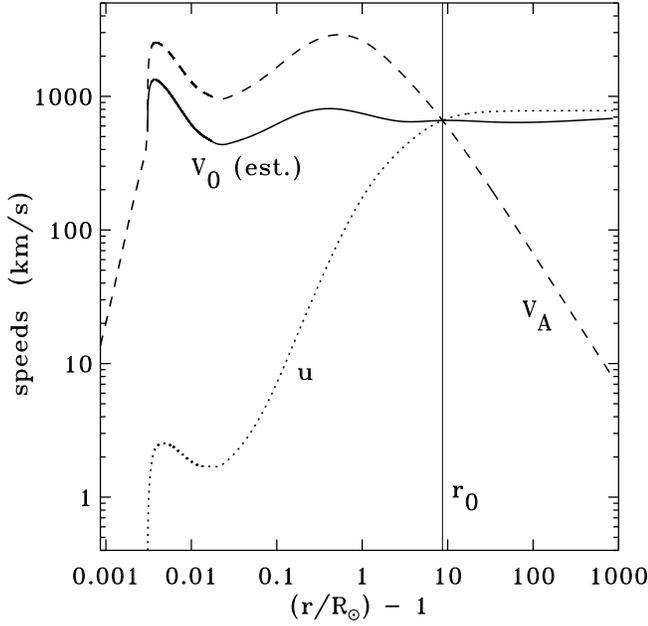}
\caption{Radial dependence of the estimated speed $V_0$ at the
Alfv\'{e}n critical point computed from Equation (\ref{eq:V0est})
(solid curve).
This is compared with the radial dependence of $V_{\rm A}$
(dashed curve) and $u$ (dotted curve) for this coronal hole model.
The estimate for $V_0$ is shown only above the sharp
chromosphere-corona transition region at
$(r/R_{\odot})-1 \approx 0.003$.}
\end{figure}

Although $r_0$ and $V_0$ can be estimated using the methods
outlined in this section, the derivatives required for computing
$\mu_0$ are likely to be less amenable to robust approximation.
Thus, when $r_0$ and $V_0$ are estimated in this way, we suggest
that Equation (\ref{eq:Winfmu}) be avoided for the high-frequency
limit of the reflection coefficient.
In this case, the bridging law given by Equation (\ref{eq:bridgem})
should be replaced by a simpler version,
\begin{equation}
  {\cal R} \, = \, {\cal R}_{\rm zero}^{1-\epsilon}
  {\cal R}_{\infty \nu}^{\epsilon}  \,\, .
  \label{eq:bridgen}
\end{equation}

\section{Results}

This section contains a detailed comparison between the numerically
computed non-WKB wave properties (i.e., exact solutions to
Equations (\ref{eq:dPsi}) and (\ref{eq:dGam})) and the results
of the approximations described in Section 2.
In Sections 3.1 and 3.2, the background coronal properties shown
are those of Cranmer \& van Ballegooijen (2005).
In Section 3.3, the approximate heating rates are compared to
those computed by Cranmer et al.\  (2007) for models of a range
of source regions of fast and slow solar wind streams.

\subsection{Reflection Coefficients}

The non-WKB reflection equations were solved numerically using an
adaptive-stepsize version of the fourth order Runge-Kutta algorithm
(e.g., Press et al.\  1992).
The baseline model consists of a grid of 350 frequency points
and 5457 radial points.
The frequency points are evenly spaced in $\log\omega$.
The minimum and maximum wave periods are 0.001 and 1000 hours,
which correspond to maximum and minimum frequencies of
$2.77 \times 10^{-1}$ and $2.77 \times 10^{-7}$ Hz respectively.
The radial grid extends from the mid-chromospheric ``merging
height'' of strong-field flux tubes
($r \approx 1.00086 \, R_{\odot}$) to a heliocentric
distance of 4 AU ($r \approx 860 \, R_{\odot}$), and the
radially varying grid separation is described by
Cranmer \& van Ballegooijen (2005).

Figure 3(a) shows the result of numerically integrating the
transport equations to compute ${\cal R}$ as a function of
radius for each ``monochromatic'' frequency in the grid.
As described in the caption, the contours in all three panels
denote constant values of ${\cal R}$ between $3 \times 10^{-5}$
and 0.9.
Figure 3(b) displays the result of using the approximations
of Sections 2.2--2.4 to compute ${\cal R}$ from
Equation (\ref{eq:bridgem}).
For Figure 3(b), the known values of $r_{0} = 9.698 \, R_{\odot}$,
$V_{0} = 660.1$ km s$^{-1}$, and $\mu_{0} = 1.184 \times 10^{-4}$
s$^{-1}$ were applied to the approximation equations.
Figure 3(c), however, shows the result of using the estimation
method of Section 2.5 to compute $r_0$ and $V_0$, and to
approximate the value of ${\cal R}$ from
Equation (\ref{eq:bridgen}).
\begin{figure}
\epsscale{1.09}
\plotone{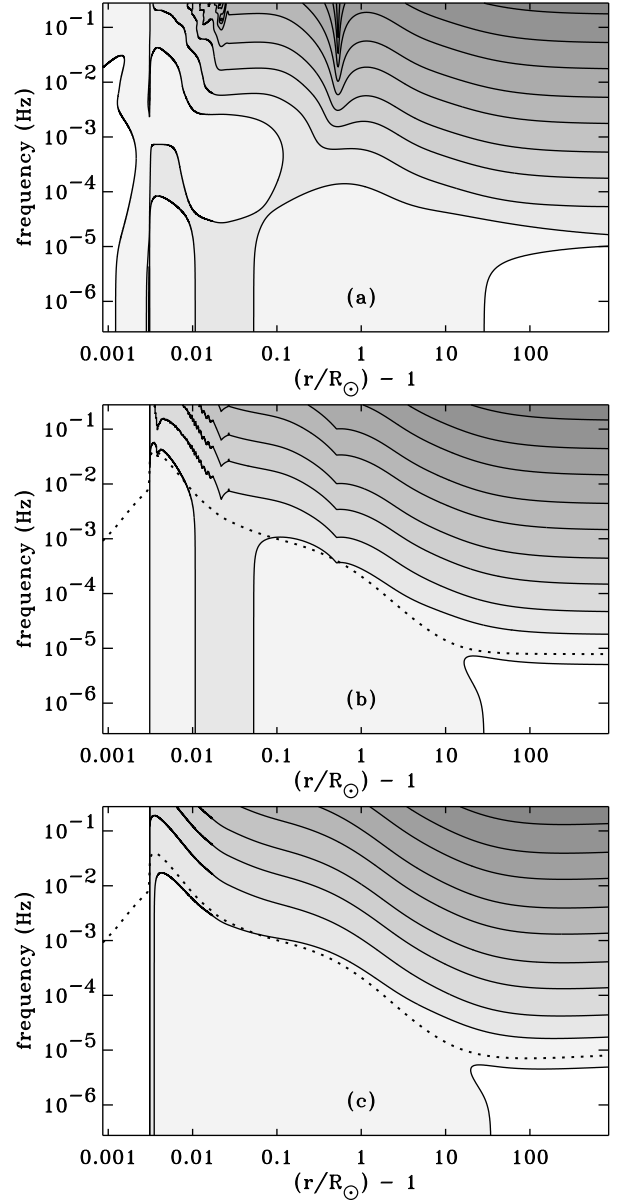}
\caption{Contour plot of the reflection coefficient
${\cal R}$ as a function of height and wave frequency.
Exact numerical results (a) are compared with results of
applying approximations of Sections 2.2--2.4 (b), as well as
the more approximate results of estimating critical
point conditions using Section 2.5 (c).
The 3 panels use identical definitions for the contour levels.
From light to dark (i.e., from bottom to top along the right-hand
side), the solid contours represent values of ${\cal R}$ of
0.9, 0.3, 0.1, 0.03, 0.01, 0.003, 0.001, $3 \times 10^{-4}$,
$10^{-4}$, and $3 \times 10^{-5}$.
Dotted lines in (b) and (c) show the bridging frequency
($\omega_{0}(r)/2\pi$) for the two approximations.}
\end{figure}

Overall, the similarities between the three panels of Figure 3
appear to outweigh the differences.
At large heights ($r \gtrsim 5 \, R_{\odot}$) the high-frequency
limiting approximations for ${\cal R}$ produce excellent agreement
with the numerical solutions.
The full radial dependence at very low frequencies ($\omega/2\pi
\lesssim 10^{-6}$ Hz) is well reproduced by the results of
Section 2.2.
There are some discrepancies between the three panels at
heights in the low corona (between 0.003 and 0.1 $R_{\odot}$)
at low and intermediate frequencies.
However, these discrepancies involve mainly the positions of
the contours labeling ${\cal R} = 0.1$ and 0.3, and the
approximations here do not show more than a factor of two
difference from the numerical solutions.
The effects of $\mu$ changing sign are apparent at heights of
about 0.02 and 0.5 $R_{\odot}$ for the highest frequencies,
but these ``spikes'' are smoothed over for frequencies lower
than about 0.01 Hz.
The average of the two high-frequency approximations
(Equation (\ref{eq:bridgem})) produces a small cusp at these
heights, which is an adequate representation of the mean radial
behavior of ${\cal R}$ at these frequencies.

In order to apply the computed reflection coefficients to a
model of turbulent dissipation and heating, one needs to
understand the properties of the continuous {\em power spectrum}
$P_{\rm A}(\omega)$ of Alfv\'{e}nic fluctuation energy in the
corona and solar wind.
In other words, one needs to know how much is contributed by
each frequency (i.e., each ``row'' in Figure 3) to the total
turbulent energy.
Although there are observational constraints on the frequency
dependence of $P_{\rm A}(\omega)$ at the solar photosphere
(Cranmer \& van Ballegooijen 2005) and there are direct in~situ
measurements at distances greater than 60 $R_{\odot}$
(Tu \& Marsch 1995; Goldstein et al.\  1997), there are no
measurements of the Alfv\'{e}nic frequency spectrum in the
extended corona and inner solar wind.
Unfortunately, the heights without measured power spectra
appear to be the most important for the specification of the
rates of heating and momentum deposition in the models.

In this paper, the frequency-weighting of the modeled reflection
coefficients will be presented for two empirically based
choices for the shape of $P_{\rm A}(\omega)$.
Figure 4 displays both spectra, each of which has been
normalized such that
\begin{equation}
  \int_{0}^{\infty} d\omega \,\, P_{\rm A}(\omega)
  \, = \, 1 \,\, .
\end{equation}
For simplicity, the normalized spectrum is assumed to retain
its shape as a function of height.
This assumption has served reasonably well in existing models
of solar wind acceleration (e.g., Cranmer et al.\  2007), but
in reality there must be some kind of spectral evolution with
increasing distance from the Sun.
A model of this evolution can be straightforwardly combined
with the approximations of Section 2, but that is beyond the
scope of this paper (see, however, Tu \& Marsch 1995;
Laitinen 2005; Verdini et al.\  2009).
\begin{figure}
\epsscale{1.17}
\plotone{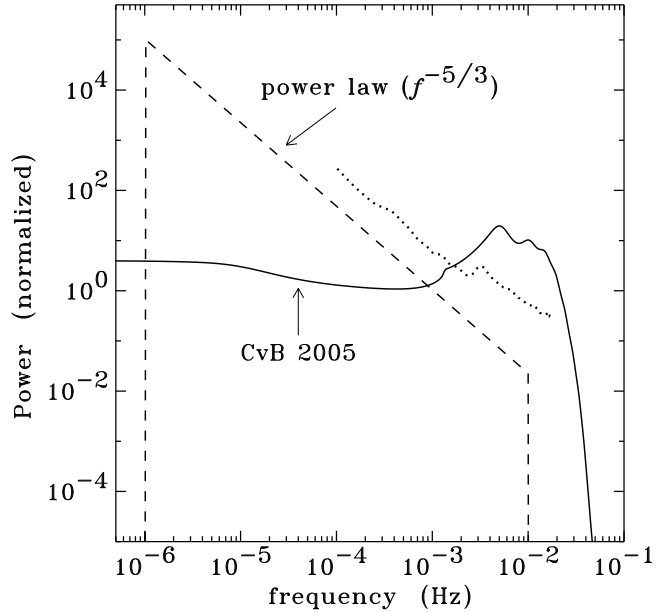}
\caption{Normalized Alfv\'{e}n wave power spectra for the
model of Cranmer \& van Ballegooijen (2005) (solid curve) and
for a Kolmogorov (1941) power law with frequency limits from
Verdini \& Velli (2007) (dashed curve).
Also shown, with arbitrary normalization, is the measured coronal
power spectrum from Tomczyk \& McIntosh (2009) (dotted curve).}
\end{figure}

The solid curve in Figure 4 shows the power spectrum of total
(kinetic plus magnetic) wave energy from the model of
Cranmer \& van Ballegooijen (2005) taken at the base of the
corona.
This model was constrained at the photospheric lower boundary
by a measured power spectrum of the kinetic motions of G-band
bright points.
These bright points represent thin magnetic flux tubes that
undergo both random walks, in response to convective
granulation, and rapid horizontal ``jumps'' that appear to be
the result of sporadic merging and fragmenting.
The spectrum at the coronal base was computed as the result
of a non-WKB model of kink-mode and Alfv\'{e}n-mode transport
in the chromosphere and transition region.
The dashed curve in Figure 4 shows a power-law spectrum
reminiscent of Kolmogorov (1941) hydrodynamic turbulence.
The lower and upper limits in frequency were obtained from the
non-WKB reflection models of Verdini \& Velli (2007), who
studied the implications of a $\omega^{-5/3}$ power law spectrum
on the coronal heating rate.

For comparison, Figure 4 also shows a measured power spectrum
of Alfv\'{e}nic fluctuations in the low corona
(Tomczyk \& McIntosh 2009) from a series of Doppler images made
by the Coronal Multi-channel Polarimeter (CoMP) at the Sacramento
Peak Observatory.
It should be noted that the coronal regions that dominated the
measured spectrum were {\em closed loops} in a coronal active
region.
Because a turbulent cascade behaves somewhat differently in open
and closed regions (e.g., Rappazzo et al.\  2008; Cranmer 2009),
there may not be a good reason to assume that this spectrum
would exist in the source regions of the solar wind.
There also may be frequency-dependent line-of-sight integration
effects that change the shape of the spectrum.\footnote{%
If the number of oscillations along the line of sight at any
one time scales with wavelength, then higher frequencies
would undergo more line-of-sight ``Doppler cancellation'' than
lower frequencies.  Thus, the intrinsic coronal spectrum would
have to be {\em flatter} than the measured spectrum.
This could imply a closer resemblance to the 
Cranmer \& van Ballegooijen (2005) spectrum than is apparent
in Figure 4.}
It is nonetheless interesting that this measured spectrum appears
to combine the overall power-law behavior of the Kolmogorov (1941)
curve with a hint of the high-frequency convective resonance
features in the Cranmer \& van Ballegooijen (2005) spectrum.

The spectrum-weighted reflection coefficient
$\langle {\cal R} \rangle$ is defined as
\begin{equation}
  \langle {\cal R} \rangle^{2} (r) \, = \,
  \frac{\int d\omega \, P_{A}(\omega) \, {\cal R}^{2}(\omega,r)}
       {\int d\omega \, P_{A}(\omega)}  \,\, ,
  \label{eq:Rweight}
\end{equation}
where the square of ${\cal R}$ is used because the power spectrum
is an energy density quantity and ${\cal R}$ is a ratio of
amplitudes.
Equation (\ref{eq:Rweight}) was used in the solar wind
acceleration models of Cranmer et al.\  (2007).
A slightly different technique was applied by
Cranmer \& van Ballegooijen (2005), who performed the weighting
on the energy densities of the $z_{+}$ and $z_{-}$ fluctuations
separately, and then computed their ratio afterwards.
The resulting heating rates from the two techniques are not
significantly different from one another.

In Figure 5, the radial dependences of the weighted reflection
coefficients $\langle {\cal R} \rangle$ are shown for the
exact and approximate calculations.
Note that ${\cal R}$ (for all frequencies) is set to 1 at heights
below the TR.
This is obviously not an exact representation of the numerically
integrated value, but---as can be seen below in Figure 6---the
resulting heating rate is insensitive to these differences.
\begin{figure}
\epsscale{1.11}
\plotone{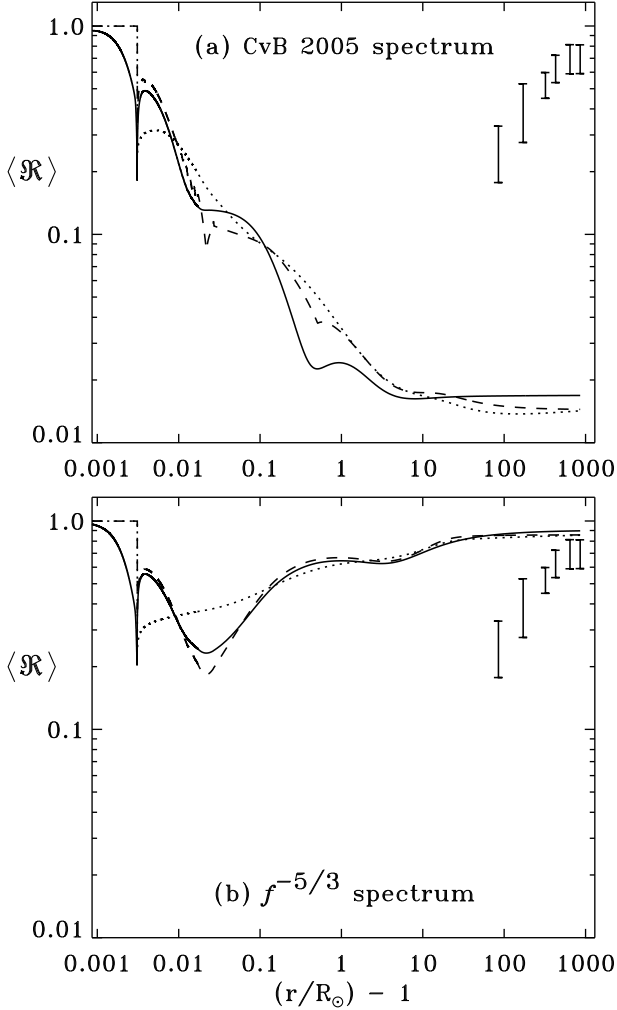}
\caption{Weighted reflection coefficients, shown as a function
of radial distance, for (a) the Cranmer \& van Ballegooijen (2005)
spectrum and (b) the truncated Kolmogorov (1941) spectrum.
In each panel, exact numerical results (solid curves) are
compared with approximate results computed with known $r_0$ and
$V_0$ (dashed curves) and the approximate results computed with
local estimates for $r_0$ and $V_0$ from Section 2.5 (dotted curves).
Measured ranges for fast-wind reflection coefficients from
{\em Helios} and {\em Ulysses} are shown as individual error bars.}
\end{figure}

Because the Kolmogorov spectrum is dominated by the lowest
frequencies, the reflection is dominated by the zero-frequency
limit of Section 2.2.
Thus, the comparison in Figure 5(b) between the exact result
and the approximation computed with known values of $r_0$ and
$V_0$ shows nearly exact agreement.
The curves in Figure 5(a) are dominated by higher frequencies
and show more of a relative discrepancy between the exact and
approximate expressions.
For both panels, the reflection coefficients
$\langle {\cal R} \rangle$ computed with the estimated values of
$r_0$ and $V_0$ (i.e., the dotted curves) show a significantly
more ``smoothed out'' radial dependence than the other two
sets of curves.
This gives rise to smoother radial variations in the heating rates.
Although this smoothing represents a departure from the exact
non-WKB results, it is probably not a major problem.
Complete models of coronal heating and solar wind acceleration
must also contain {\em heat conduction,} and
no matter what fine radial structure may exist in the heating
rate itself, the existence of conduction gives rise to a similar
kind of smearing of the thermal energy along the magnetic field.

Figure 5 also shows {\em Helios} and {\em Ulysses} 
measurements in the fast solar wind that have been processed
from the Els\"{a}sser energy densities presented by
Bavassano et al.\  (2000).
The upper and lower limits of the error bars reflect the spread
in individual data points, each of which represented one-hour
datasets.
The frequency range of the fluctuations sampled by these
measurements spanned only about an order of magnitude, from
$3 \times 10^{-4}$ Hz to $4 \times 10^{-3}$ Hz.
As can be seen in Figure 4, this range of frequencies is where
the two theoretical spectra have comparable power to one another.
It is the frequencies {\em outside} this narrow range that
give rise to the significant differences between the modeled
sets of curves in Figures 5(a) and 5(b).
Contributions from higher frequencies drive down the reflection
coefficient in Figure 5(a), and contributions from lower
frequencies drive it up in Figure 5(b).
Thus, it is not surprising that the measurements of power at
intermediate frequencies give reflection coefficients that fall
between the two sets of curves.

\subsection{Turbulent Heating Rates}

The adopted phenomenological rate of coronal heating is an
expression for the total energy flux that cascades from large
to small scales.
It is constrained by the properties of the fluctuations at the
largest scales, and it does not specify the exact kinetic means
of dissipation once the energy reaches the smallest scales.
Dimensionally, this is similar to the rate of cascading energy
flux derived by von K\'{a}rm\'{a}n \& Howarth (1938) for
isotropic hydrodynamic turbulence.
The full form, which takes into account various MHD effects, is
\begin{equation}
  Q_{\rm turb} \, = \, \rho \, {\cal E}_{\rm turb} \,
  \frac{Z_{-}^{2} Z_{+} + Z_{+}^{2} Z_{-}}{4 L_{\perp}}
  \label{eq:Qturb}
\end{equation}
(Hossain et al.\  1995; Zhou \& Matthaeus 1990;
Matthaeus et al.\  1999; Dmitruk et al.\  2001, 2002;
Breech et al.\  2008).
The individual components of Equation (\ref{eq:Qturb}) are
described below.

The absolute values of the spectrum-weighted Els\"{a}sser
variables are denoted $Z_{+}$ and $Z_{-}$.
Strictly speaking, in a model of non-WKB wave reflection the
power in both components can be specified only by combining the
derived reflection coefficient ${\cal R}$ with a specification
of the absolute power spectrum of the fluctuation energy.
However, the assumption that the {\em total} wave power varies
in accord with straightforward wave action conservation has been
shown to be reasonable, even in interplanetary space where
${\cal R}$ is not small (Zank et al.\  1996;
Cranmer \& van Ballegooijen 2005).
The examples shown below utilize this assumption in order to
compute $Z_{+}$ and $Z_{-}$.

Under wave action conservation, the product of a (modified)
energy flux and the transverse area of the flux tube remains
constant as a function of radial distance.
For dispersionless Alfv\'{e}n waves, this is equivalent to
maintaining a constant energy flux per unit magnetic field
strength, or
\begin{equation}
  \frac{F}{B_0} \, = \,
  \frac{(u + V_{\rm A})^{2} U_{\rm A}}{V_{\rm A} B_{0}}
  \, = \, \mbox{constant ,}
  \label{eq:fob}
\end{equation}
where the wave energy density $U_{\rm A} = \rho v_{\perp}^{2}$
(see, e.g., Jacques 1977).
Measurements of $v_{\perp}$ can be made by analyzing the
nonthermal Doppler broadening of coronal emission lines.
At $r \approx 1.1 \, R_{\odot}$, measurements made by the
SUMER instrument on {\em SOHO} corresponded to a range of
perpendicular velocity amplitudes $v_{\perp} \approx 50$--60
km s$^{-1}$ (Banerjee et al.\  1998; Landi \& Cranmer 2009).
Using the plasma parameters from the coronal hole model
discussed above, these amplitudes are consistent with a
range of values for $F/B_{0}$ between about $5 \times 10^{4}$
and $8 \times 10^{4}$ erg cm$^{-2}$ s$^{-1}$ G$^{-1}$.
At $r \approx 1.5 \, R_{\odot}$, lower and upper limit
measurements made by the UVCS instrument on {\em SOHO} gave
a range of amplitudes $v_{\perp} \approx 100$--140 km s$^{-1}$
(Esser et al.\  1999).
These are consistent with a range of $F/B_{0}$ values similar
to those at the lower height: $6 \times 10^{4}$ to
$10^{5}$ erg cm$^{-2}$ s$^{-1}$ G$^{-1}$.

The empirically constrained models of
Cranmer \& van Ballegooijen (2005) were used as another way
to put limits on the most likely wave amplitudes in the corona.
These models included wave dissipation due to the presence of
turbulence, so the ``constant'' $F/B_{0}$ actually decreases
slightly with increasing distance.
In those models, the numerical values of $F/B_{0}$ ranged
between about $2 \times 10^{4}$ and $9 \times 10^{4}$
erg cm$^{-2}$ s$^{-1}$ G$^{-1}$.
For the remainder of this paper, an intermediate constant
value of $5 \times 10^{4}$ erg cm$^{-2}$ s$^{-1}$ G$^{-1}$
is assumed.
With that, the full radial dependence of $U_{\rm A}$ can be
computed, and the spectrum-averaged Els\"{a}sser amplitudes
can be determined from
\begin{equation}
  Z_{-} \, = \, \sqrt{\frac{4 U_{\rm A}}{\rho (1 +
  \langle {\cal R} \rangle^{2} )}}
  \,\, , \,\,\,\,
  Z_{+} \, = \, \langle {\cal R} \rangle Z_{-}
\end{equation}
for use in Equation (\ref{eq:Qturb}).

The perpendicular length scale $L_{\perp}$ is an effective
transverse correlation length of the turbulence for the
largest eddies.
The models presented here use a standard assumption that the
correlation length scales with the transverse width of the
magnetic flux tube; i.e., that $L_{\perp} \propto B_{0}^{-1/2}$
(see also Hollweg 1986).
Cranmer \& van Ballegooijen (2005) found that the following
normalization produced a reasonable combination of heating
and wave dissipation:
\begin{equation}
  L_{\perp}(r) \, \approx \,
  \frac{11.55}{\sqrt{B_{0}(r)}} \,\, \mbox{Mm ,}
  \label{eq:Lperpdef}
\end{equation}
where $B_0$ is measured in Gauss.
However, the self-consistent models of Cranmer et al.\  (2007)
worked best by reducing the normalization by about a factor of
four; i.e., by replacing the factor of 11.55 in the above by
the smaller value 2.876.
With no convincing evidence to choose between these two values,
the example models shown below use a compromise value of
6.0 Mm for this normalization.

The quantity ${\cal E}_{\rm turb}$ in Equation~(\ref{eq:Qturb})
is an efficiency factor that attempts to account for regions
where the turbulent cascade may not have time to develop before
the fluctuations are carried away by the wind.
Cranmer et al.\  (2007) estimated this efficiency factor to
scale as
\begin{equation}
  {\cal E}_{\rm turb} \, = \, \frac{1}{1 +
  ( t_{\rm eddy} / t_{\rm ref} )^{n}}  \,\, ,
  \label{eq:Eturb}
\end{equation}
where the two timescales above are $t_{\rm eddy}$, a nonlinear
eddy cascade time, and $t_{\rm ref}$, a timescale for large-scale
Alfv\'{e}n wave reflection.
The value $n=1$ was chosen for the exponent based on a range of
analytic and numerical turbulence models (Pouquet et al.\  1976;
Dobrowolny et al.\  1980; Matthaeus \& Zhou 1989;
Dmitruk \& Matthaeus 2003; Oughton et al.\  2006).
Equation (\ref{eq:Eturb}) quenches the turbulent heating when
$t_{\rm eddy} \gg t_{\rm ref}$, i.e., when the Alfv\'{e}n waves
want to propagate away much faster than the cascade can proceed
at a given location.
Cranmer et al.\  (2007) defined the reflection time as
$t_{\rm ref} = 1/ |\nabla \cdot {\bf V}_{A}|$.
For the purposes of this paper, the simpler approximation
$t_{\rm ref} = \nu^{-1}$ was found to give an equivalent result
(see Equation (\ref{eq:nu})).
This latter definition is used in the results presented below.
The eddy cascade time is given by
\begin{equation}
  t_{\rm eddy} \, = \, \frac{L_{\perp} \sqrt{3\pi}}
  {(1 + M_{A}) \, v_{\perp}}  \,\, ,
\end{equation}
where the Alfv\'{e}n Mach number $M_{A} = u/V_A$ and the
numerical factor of $3\pi$ comes from the normalization
of an assumed shape of the turbulence spectrum (see
Appendix C of Cranmer \& van Ballegooijen 2005).

Figure 6 shows the computed heating rates for the various
spectrum-weighted cases discussed above.
Rather than plot $Q_{\rm turb}$ itself, Figure 6 shows the
heating rate per unit mass $Q_{\rm turb}/\rho$ in order to more
clearly indicate which heights receive the most heating on a
particle-by-particle basis.
As in Figure 5, the differences between the exact and approximate
non-WKB results are relatively minor.
\begin{figure}
\epsscale{1.15}
\plotone{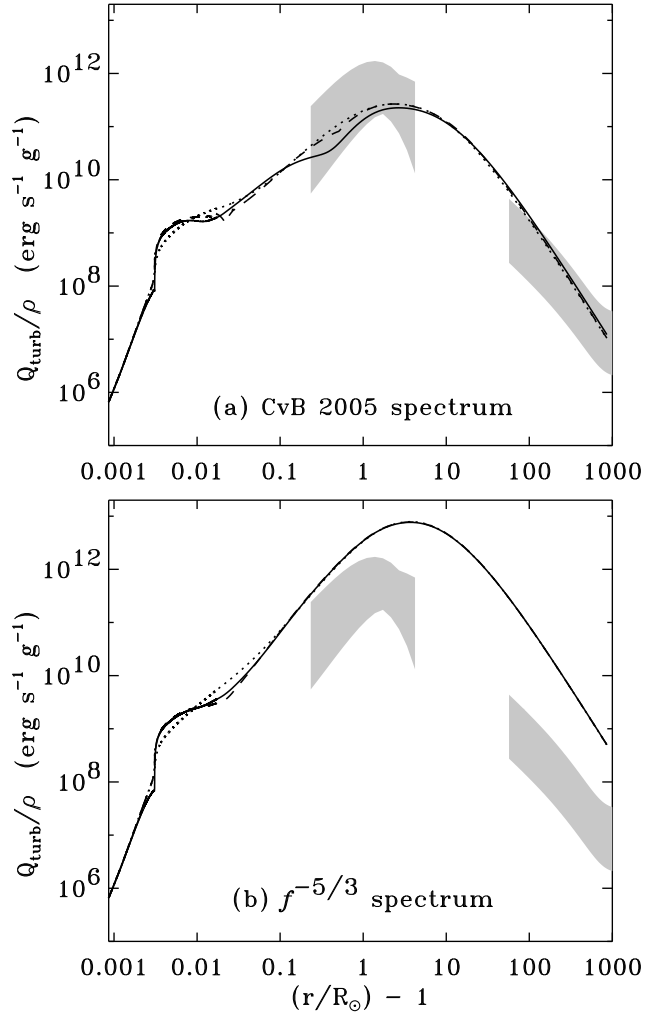}
\caption{Turbulent heating rate per unit mass
($Q_{\rm turb}/ \rho$) for the coronal hole models.
The curves shown in panels (a) and (b), and the line plotting
styles in each panel, are the same as in Figure 5.
The gray regions show empirically constrained heating rates
(see the text).}
\end{figure}

Each panel of Figure 6 also shows observational constraints
on the heating rate in the fast solar wind.
The gray region in the corona (i.e., at heights between 0.2 and 6
$R_{\odot}$ above the surface) delineates the lower and upper
bounds on a set of parameterized heating rates taken from a
range of papers (Wang 1994; Hansteen \& Leer 1995;
Allen et al.\  1998; Cranmer et al.\  2007).
These are heating rates that are {\em required} for these
one-dimensional coronal models to be able to produce realistic
fast solar wind conditions.
Figure 2(b) of Cranmer (2004) illustrates some of the individual
heating functions that were collected into this set.
The gray region in interplanetary space (i.e., at heights greater
than $\sim$60 $R_{\odot}$) illustrates the range of
total ($Q_{p} + Q_{e}$) heating rates determined empirically
by Cranmer et al.\  (2009) from {\em Helios} and {\em Ulysses}
plasma measurements in the fast wind.

Note that the heating rates computed from the
high-frequency-dominated spectrum of
Cranmer \& van Ballegooijen (2005)---shown in Figure 6(a)---tend
to be in better agreement with the empirical constraints than
do the heating rates computed from the low-frequency-dominated
spectrum, in Figure 6(b).
This is consistent with the results of Cranmer et al.\  (2007),
in which a similar high-frequency-dominated spectrum was used
to produce reasonably successful models of both fast and slow
solar wind streams.
Note also that both sets of model results in Figure 6 overestimate
the measured heating in interplanetary space.
This is likely to be the result of modeling the radial dependence
of $Z_{\pm}$ with wave action conservation.
In the models of Cranmer \& van Ballegooijen (2005), the turbulent
dissipation produced a factor of 4 reduction in $v_{\perp}$ at
1 AU, from 144 km s$^{-1}$ (undamped) to 35.2 km s$^{-1}$ (damped).
This would give about a factor of $4^{3/2} = 8$ decrease in the
heating rate far from the Sun.
Inserting this into the present models would push the curves in
Figure 6(a) to the bottom of the empirical range, and it would push
the curves in Figure 6(b) down to the top of the empirical range.

\subsection{Fast and Slow Wind Source Regions}

Although the numerical examples shown above were computed for
the fast solar wind that emerges from a polar coronal hole, the
approximations developed in this paper are not meant to be limited
to only that type of solar wind.
Figure 7 shows heating rates computed not only for coronal holes,
but also source regions of {\em slow wind} at solar minimum (i.e.,
the legs of quiescent equatorial streamers) and at solar maximum
(i.e., active regions).
These models are self-consistent numerical calculations of the
non-WKB reflection, turbulent heating, and wind acceleration that
were described by Cranmer et al.\  (2007).
The quiescent streamer model corresponds to the ``last'' open
field line that originates at a colatitude of {29.7\arcdeg} in
the two-dimensional axisymmetric magnetic field model of
Banaszkiewicz et al.\  (1998).
The active region model has an added component to its magnetic field
strength that was parameterized by Cranmer et al.\  (2007) as
having an exponential height dependence of
$B_{\rm A} e^{-(r-R_{\odot})/h}$, where $B_{\rm A} = 50$ G
and $h = 0.07 \, R_{\odot}$.
\begin{figure}
\epsscale{1.09}
\plotone{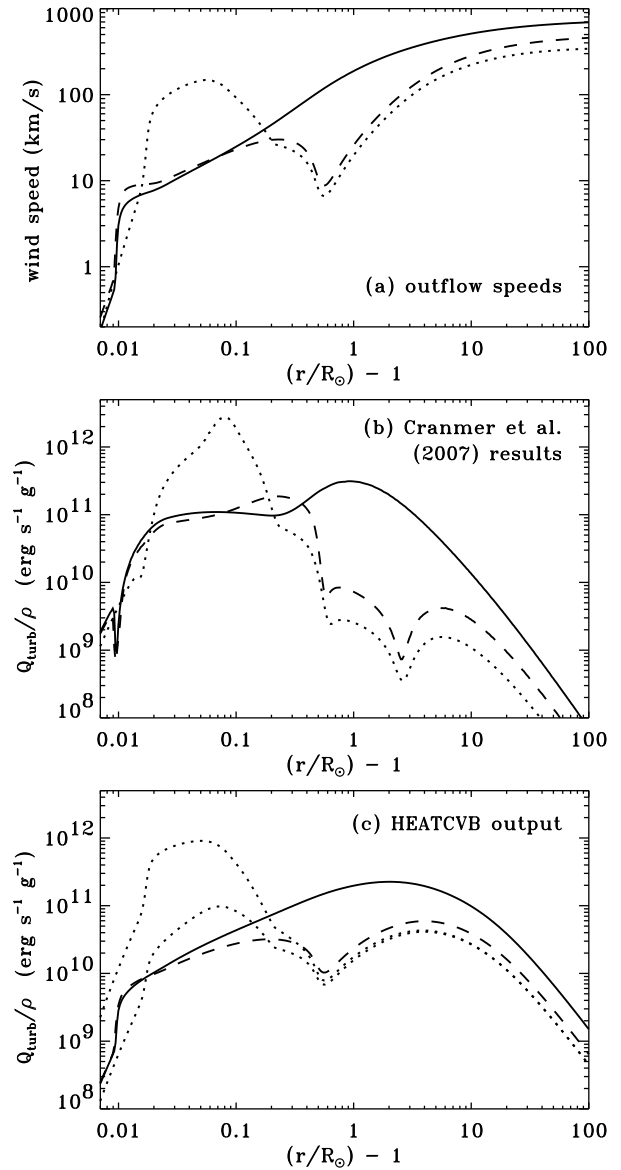}
\caption{Comparisons of plasma parameters for source regions of
fast and slow solar wind: i.e., polar coronal holes (solid
curves), quiescent equatorial streamers (dashed curves), and
active region field lines connected to the solar wind (dotted
curves).
Numerical results for (a) outflow velocity, and (b) heating
rate per unit mass from Cranmer et al.\  (2007) are compared
with heating rates computed using the HEATCVB code (c).}
\end{figure}

Figure 7(a) shows the radial dependence of the wind speed $u$
for these three models.
Note that the active region model has an outflow speed of order
100 km s$^{-1}$ in the low corona.
This appears to correspond with recent measurements of active
region outflows of similar magnitude made by the Extreme-ultraviolet
Imaging Spectrometer (EIS) on {\em Hinode} (see, e.g.,
Harra et al.\  2008).
Figure 7(b) presents the heating rates per unit mass
($Q_{\rm turb}/\rho$) that were computed self-consistently in
the Cranmer et al.\  (2007) models.
Figure 7(c) shows the heating rates as computed under the
approximations of this paper.

For the active region slow wind model, two curves are shown in
Figure 7(c).
The lower curve was computed using the standard form for the
dimensionless turbulent efficiency factor ${\cal E}_{\rm turb}$
given by Equation (\ref{eq:Eturb}).
The upper curve was computed by not using this efficiency factor
(i.e., by assuming ${\cal E}_{\rm turb} = 1$).
Clearly, the strong peak in the heating rate that is seen in
Figure 7(b) is reproduced much more satisfactorily when the
efficiency factor is not used.
The other two models---for the coronal hole and streamer
cases---did not exhibit as strong a relative difference in
the heating rate when ${\cal E}_{\rm turb}$ was set to unity.
It should be noted that this factor was not used in the earliest
published versions of Equation (\ref{eq:Qturb}) (e.g.,
Hossain et al.\  1995), nor is it used in other more recent
studies (e.g., Chandran \& Hollweg 2009).
Thus, this term should still be viewed as an approximate attempt
to take account of ``wave escape'' effects and not the final
word in that story.
Future improvements in modeling the turbulence-driven solar wind
should involve finding a more robust way of expressing how
turbulence is quenched in regions where waves escape more rapidly
than they can cascade and dissipate.

\section{Numerical Code}

A brief FORTRAN~77 subroutine called HEATCVB has been
developed to implement the approximations and other
expressions given in Sections 2 and 3.
The source code is included with this paper as online-only material,
and it is also provided, with updates as needed, on the author's
web page.\footnote{http://www.cfa.harvard.edu/$\sim$scranmer/}
There are only four required inputs to this subroutine:
the radial distance $r$, the wind speed $u$, the mass density
$\rho$, and the magnetic field strength $|B_{0}|$.
The primary output of the subroutine is the heating rate
$Q_{\rm turb}$ at the location defined by the input parameters.
All quantities are assumed to be in cgs/Gaussian units.

In addition to the four required input parameters, the
HEATCVB subroutine has three {\em optional} input parameters:
the radius of the Alfv\'{e}n critical point $r_0$,
the speed at this location (i.e.,
$V_{0} = u(r_{0}) = V_{\rm A}(r_{0})$),
and the local Alfv\'{e}n wave velocity amplitude $v_{\perp}$.
If the input values of $r_0$ or $V_0$ are less than or equal to
zero, the code will recompute them using the estimation method
described in Section 2.5.
If the input value of $v_{\perp}$ is less than or equal to zero,
it will be recomputed using wave action conservation (i.e.,
Equation (\ref{eq:fob})).
This approximation appears to be reasonably valid in the corona
(where most of the heating occurs), but it overestimates the
Alfv\'{e}n wave amplitudes in interplanetary space.

Another derived quantity that is computed and output by HEATCVB
is the local turbulent dissipation rate of the waves,
$\gamma = Q_{\rm turb} / (2 U_{\rm A})$.
This quantity is specified in units of s$^{-1}$ and can be used
when solving for {\em departures} from wave action conservation.
The factor of 2 in the above definition assumes the standard
convention that $\gamma$ is the damping rate for the wave
amplitude.
If instead the damping rate for wave energy is required,
$\gamma = Q_{\rm turb} / U_{\rm A}$.
Cranmer \& van Ballegooijen (2005) showed how this kind of wave
dissipation is necessary for agreement between measurements of
Alfv\'{e}n wave amplitudes in the corona and in the heliosphere
(see also Roberts 1989; Verdini \& Velli 2007).

For simplicity, the HEATCVB code does not utilize the $\mu$
gradients described in Section 2.3.
Thus, the bridging between the low-frequency and high-frequency
limits for ${\cal R}$ is computed using Equation (\ref{eq:bridgen})
instead of Equation (\ref{eq:bridgem}).
The weighting of ${\cal R}$ over the Alfv\'{e}nic power spectrum
is performed using the high-frequency-dominated spectrum of
Cranmer \& van Ballegooijen (2005).
As described above, the use of this particular choice of the
spectrum worked well for the self-consistent solar wind models
of Cranmer et al.\  (2007).
The integration over the normalized power spectrum is discretized
into bins separated by factors of two in frequency space (i.e.,
``octaves'').
The wave power in each bin $i$ is summed into weighting factors
$f_i$, such that Equation (\ref{eq:Rweight}) is approximated by
\begin{equation}
  \langle {\cal R} \rangle^{2} \, \approx \,
  \sum_{i} f_{i} \, {\cal R}^{2}(\omega_{i}) \,\, .
\end{equation}
To make the above compatible with Equation (\ref{eq:Rweight}), the
weights $f_i$ are defined such that they sum to 1.
The HEATCVB code uses 17 bins that span more than four orders of
magnitude in frequency space.
The code also contains several checks to ensure that the input
parameters are within reasonable bounds for the solar atmosphere
and solar wind.
All major steps in the algorithm are described in comments within
the source code.

The HEATCVB routine contains several tests to evaluate
whether a full calculation of the non-WKB reflection is warranted.
Whenever $u \leq 0$, the code avoids this calculation and sets
$\langle {\cal R} \rangle = 1$.
The general assumption in this case is that the local plasma is
in a closed-field region.
The condition $u=0$ could correspond to a hydrostatic coronal
streamer, and $u < 0$ could signal the existence of transient
downflows similar to those observed with visible-light
coronagraphs (Wang et al.\  1999).
In flux tubes where both footpoints are anchored to the solar
surface, Alfv\'{e}n waves are believed to propagate up and down
with nearly equal intensities in the two Els\"{a}sser components
(see, e.g., Rappazzo et al.\  2008).
Also, whenever $\rho \geq 2 \times 10^{-13}$ g cm$^{-3}$ the
local plasma conditions are judged to be ``chromospheric.''
In this case, the code sets $\langle {\cal R} \rangle = 1$
because it assumes the location to be modeled is below the sharp
reflection barrier of the TR.
This density criterion may also be triggered for coronal mass
ejections (CMEs).
These regions may be similar to the above case of coronal loops,
since the field lines are closed in CME plasmoids, and the
shocked regions in front of the plasmoids are generally believed
to have both footpoints rooted to the solar surface
(e.g., Lin et al.\  2003).
However, in CMEs the assumption of time-steady wave action
conservation may not give an accurate value for the local
fluctuation amplitude $v_{\perp}$.

\section{Conclusions}

The primary aim of this paper has been to develop and test a set
of approximations to the solutions of the equations of non-WKB
Alfv\'{e}n wave reflection.
These solutions are designed to be applied to modeling the plasma
heating in open-field regions of the solar corona and the solar wind.
Two independent approximate expressions for the reflection
coefficients ${\cal R}$ were developed in the limiting cases of
extremely low wave frequencies (Section 2.2) and extremely high
wave frequencies (Section 2.3).
These, together with a technique to estimate the radius and speed
at the Alfv\'{e}n critical point, provide a robust approximation
for the necessary ingredients of the phenomenological turbulent
heating rate.
The resulting radial dependence of ${\cal R}$ and the heating rate
$Q_{\rm turb}$ were shown to agree well with exact solutions of
the non-WKB transport equations in several cases relevant to the
fast and slow solar wind.

Because the approximations presented above do not depend on
computationally intensive integrations along flux tubes, they
make it much easier to insert more realistic (wave/turbulence
heating) physics into three-dimensional models of the corona
and heliosphere.
An important first step is to perform ``testbed'' simulations for
specific time periods in which large amounts of empirical data are
available.
Such times include the Whole Sun Month in 1996 (Galvin \& Kohl 1999)
and the Whole Heliosphere Interval in 2008 (Gibson et al.\  2009).
These testbed models can help determine whether, for example,
the WTD or RLO paradigms for solar wind acceleration are better
representations of the actual physics (see Section 1).
These models will also be key to assessing and validating
real-time {\em predictive} models of the plasma conditions in
the heliosphere.

A significant remaining unknown quantity in the above modeling
methodology is the shape of the frequency spectrum of Alfv\'{e}n
waves in the corona.
The HEATCVB code utilizes an empirically constrained spectrum
from the work of Cranmer \& van Ballegooijen (2005).
This particular form of the spectrum was shown by
Cranmer et al.\  (2007) to produce reasonably successful
self-consistent models of both fast and slow solar wind
streams, so it appears to be a good first approximation.
However, this can be replaced easily with other forms of the
power spectrum if better constraints become available.
It is also likely that the shape of the power spectrum depends
on radial distance and cannot be specified universally for all
values of $r$ (e.g., Verdini et al.\  2009).
Once such a radial dependence is specified, however, the
approximations developed in this paper can still be used to
estimate the turbulent heating at any location along the open
field lines.

In order to make further progress, it will be important to
include other physical processes in WTD-type models of coronal
heating and solar wind acceleration.
In addition to non-WKB reflection, there are other ways that
the energy in outward propagating waves can be tapped to produce
inward propagating waves.
Large-scale shear motions in the heliosphere have been suggested
as a source of kinetic energy that could go into reducing the
overall cross helicity of the turbulence (e.g.,
Roberts et al.\  1992; Zank et al.\  1996; Matthaeus et al.\  2004;
Breech et al.\  2008; Usmanov et al.\  2009).
Kinetic instabilities in collisionless regions of the corona
and solar wind can give rise to the local generation of
high-frequency inward propagating waves (Isenberg 2001;
Isenberg et al.\  2009).
Also, outward propagating Alfv\'{e}n waves of sufficient amplitude
may become unstable to nonlinear processes---such as ``parametric
decay''---which give rise to enhanced inward propagating wave
activity (e.g., Goldstein 1978; Jayanti \& Hollweg 1993;
Lau \& Siregar 1996; Ofman \& Davila 1998;
Suzuki \& Inutsuka 2006; Hollweg \& Isenberg 2007).

Although the phenomenological heating rate given in
Equation (\ref{eq:Qturb}) was motivated by the results of
many analytic and numerical studies, there is still some
disagreement about its robustness and applicability.
Chandran et al.\  (2009) found that the exact dependence of
the dissipation rate on $Z_{+}$ and $Z_{-}$ may depend on the
specific generation mechanism(s) of inward propagating waves.
Also, the evolution of the correlation length $L_{\perp}$ with
radial distance should be coupled to the overall evolution of
$Z_{\pm}$ and not simply scaled with the flux-tube width as in
Equation (\ref{eq:Lperpdef}) (see, e.g., Matthaeus et al.\  1999;
Breech et al.\  2009).

Finally, it will be important for future models to take account
of the {\em multi-fluid} nature of coronal heating and solar
wind acceleration.
A large-scale description of the energy flux injected into the
turbulent cascade is needed in order to model its eventual
kinetic dissipation (and the subsequent preferential
energization of electrons, protons, and heavy ions).
Even in the strongly collisional ``low corona,'' there are
likely to be macroscopic dynamical consequences of the
partitioning of energy between protons and electrons
(see, e.g., Hansteen \& Leer 1995; Endeve et al.\  2004).
Non-WKB Alfv\'{e}n wave reflection also affects the energy
and momentum coupling between protons and other ions in the
solar wind (Li \& Li 2007, 2008).
Including these effects can lead to concrete predictions for
measurements to be made by space missions such as
{\em Solar Probe} (McComas et al.\  2007) and {\em Solar
Orbiter} (Marsden \& Fleck 2007), as well as next-generation
ultraviolet coronagraph spectroscopy that could follow up on
the successes of the UVCS instrument on {\em SOHO}
(Kohl et al.\  2006).

\acknowledgments

I gratefully acknowledge Adriaan van Ballegooijen, William
Matthaeus, and Ofer Cohen for many valuable discussions, as well
as Steve Tomczyk and Scott McIntosh for making available a
tabulated version of their CoMP power spectrum.
I also thank the anonymous referee for constructive suggestions
that have improved this paper.
This work was supported by the National Aeronautics and Space
Administration (NASA) under grants {NNG\-04\-GE77G} and
{NNX\-09\-AB27G} to the Smithsonian Astrophysical Observatory.

\end{document}